# Measuring the response of gold prices to uncertainty: An analysis beyond the mean[1]

Jamal Bouoiyour
University of Pau, France.
Email: jamal.bouoiyour@univ-pau.fr

Refk Selmi
University of Pau, France.
Email: refk.selmi@univ-pau.fr

Mark E. Wohar
Corresponding author. College of Business Administration, University of Nebraska at Omaha, 6708 Pine Street, Omaha, NE 68182, USA; School of Business and Economics, Loughborough University, Leicestershire, LE11 3TU, UK.
Email: mwohar@unomaha.edu

**Abstract:** This paper provides an innovative perspective on the role of gold as a hedge and safe haven. We use a quantile-on-quantile regression approach to capture the dependence structure between gold returns and changes in uncertainty under different gold market conditions, while considering the nuances of uncertainty levels. To capture the core uncertainty effects on gold returns, a dynamic factor model is used. This technique allows summarizing the impact of six different indexes (namely economic, macroeconomic, microeconomic, monetary policy, financial and political uncertainties) within one aggregate measure of uncertainty. In doing so, we show that the gold's role as a hedge and safe haven cannot be assumed to hold at all times. This ability seems to be sensitive to the gold's various market states (bearish, normal or bullish) and to whether the uncertainty is low, middle or high. Interestingly, we find a positive and strong relationship between gold returns and the uncertainty composite indicator when the uncertainty attains its highest level and under normal gold market scenario. This suggests that holding a diversified portfolio composed of gold could help protecting against exposure to uncertain risks.

---

[1] ***Acknowledgement:*** The authors would like to thank the editor Sushanta Mallick and the two anonymous Reviewers for helpful and insightful comments and suggestions on an earlier version of this article.

**Highlights**

- We examine the response of gold returns to different uncertainty indicators.
- We use a quantile-on-quantile regression model.
- We develop an uncertainty composite indicator based on six uncertainty factors.
- The hedge and safe haven ability of gold is conditional on gold market states.
- The gold's role as a hedge and safe haven depends on the nuances of uncertainty.

# 1. Introduction

The gold's hedge and safe haven property is one of the most investigated issues in finance. Gold is widely viewed as a safe haven[2] when other asset classes are very volatile. Previous empirical studies documented that the linkage between gold and other assets changes noticeably in times of market stress or turmoil (Hartmann et al. 2004; Baur and McDermott, 2010; Baur and Lucey, 2010; Wang et al., 2011; Miyazaki et al., 2012; Ciner et al. 2013; Mensi et al., 2013; Reboredo, 2013; Wang and Chueh, 2013; Arouri et al., 2015; Bampinas and Panagiotidi, 2015; Beckmann et al., 2015; Nguyen et al., 2016; Van Hoang et al., 2016, among others). Although the empirical literature on the dependence between gold prices and other assets is increasing remarkably, Beckmann et al. (2017) suggested that the traditional and the well-known view on hedge and safe haven properties of gold can be misleading and that it seems more relevant to directly assess the dependence between gold and uncertainty.

There is a limited strand of literature that seeks to investigate the effects of uncertainty on gold price dynamics. All of them focus on individual uncertainty proxies. Jones and Sackley (2016) investigated the dependence between gold price dynamics and the economic policy uncertainty, and found that a great uncertainty leads to an increase in gold prices. Likewise, Gao and Zang (2016) examined the impacts of economic policy uncertainty on the relationship between the UK stock market and gold market. They showed that uncertain economic policies prompt a moderate correlation, while a strong correlation is associated with certain economic policies. Balcilar et al. (2016) tested the role of economic, macroeconomic and financial uncertainty in determining gold returns and volatility, and found that macroeconomic and financial uncertainties play a vital role in explaining gold price dynamics. Nevertheless, there is no significant effect of the economic-policy uncertainty on gold returns. Moreover, Gospodinov and Jammali (2016) evaluated the effects of monetary policy uncertainty on commodity prices. They argued that the uncertainty associated with negative monetary policy shocks drops commodity prices, whereas the uncertainty associated with positive monetary policy shocks affects varyingly (i.e., negatively and positively depending to the considered commodity) the prices of commodities. Furthermore, the association between asset market developments and politics has also been largely investigated in the literature dealing with the implications of political cycles as well as the political orientation of the government for asset returns (see, for instance, Döpke and Pierdzioch, 2006; Bohl et al. 2009; Gupta et al., 2017; Hou et al. 2017, etc.). Hou et al. (2017) investigated the impact of political uncertainty surrounding the U.S. presidential elections on the prices of 87 commodities. They found that the political uncertainty exerts a statistically negative influence on commodity prices (in particular, gold prices). Interestingly, the empirical works that compare the impacts of several uncertainty measures on gold returns are very scarce,[3] and showed wide differences in their effect on gold price dynamics (for instance, Balcilar et al., 2016; Beckmann et al., 2017).

---
[2] A safe haven is an investment that is expected to retain its value or even rise in times of turmoil.

Our research complements prior empirical studies by investigating whether gold satisfies hedge and safe haven properties under different gold market scenarios and diverse uncertainty episodes. It is evident that both time and frequency are prominent for gold price dynamics as gold has experienced a sharp evolution in recent years (Fang et al., 2018). Nevertheless, several works have differentiated the short-term and long-term correlations between gold and other assets (see, for instance, Wang et al., 2011 ; Wang and Chueh, 2013 ; Beckmann et al., 2015 ; Nguyen et al., 2016 ; Gao and Zang, 2016). Also, some studies have tested whether gold provides the ability of hedging against inflation by discriminating between short-run and long-run dynamics (see, for example, Van Hoang et al., 2016). The novelty of our study lies in the analysis of the entire dependence structure of the quantile of gold returns and that of different uncertainty indicators, thereby extending the quantile regression to a quantile-on-quantile regression (QQR). This method provides a measure of average dependence as well as of upper and lower tail dependence, conditional on dissimilar of gold market conditions and on whether the uncertainty is low, middle or high.

---

[3] It must be stressed at this stage that all of the aforementioned studies concentrate on a particular aspect of uncertainty while we conduct a more comprehensive perspective. In the present research, we differentiate between distinct types of uncertainty, while the papers on the central issue focus on some uncertainty measures (with large extent, the macroeconomic uncertainty and the economic policy uncertainty). Besides, no many empirical studies consider the linkage between uncertainty and variables other than the economic activity.

Moreover, this study proposes an uncertainty composite indicator aimed at synthesizing the impacts of six uncertainty indexes (namely, economic, macroeconomic, microeconomic, monetary policy, financial and political uncertainties) within one aggregate measure of uncertainty. To do so, we use a dynamic factor model (DFM) proposed by Doz et al. (2012). The new uncertainty composite indicator takes into account the elemental uncertainty dynamics, and then allows capturing the core effects of uncertainty on gold returns, which are not specific to a particular measure of this phenomenon. To the best of our knowledge, this is the first paper to examine the ability of gold to hedge against the core uncertainty effects. There has been very little research reported on the development of an aggregate measure of the uncertainty using different indicators (Haddow et al., 2013 ; Charles et al., 2017). However, no studies were found on the effect of a composite uncertainty index on gold returns. In short, the use of a dynamic factor model and a quantile-on-quantile regression give investors a much broader and more accurate picture than looking at just how the relationship between gold and other assets vary over time.

Our findings reveal that gold would serve as a hedge and a safe haven in uncertain times. But this property depends on gold market circumstances, the nuances of uncertainty levels as well as the measures of uncertainty used. Our results also underscore the usefulness of the uncertainty composite indicator.

The remainder of the paper is organized as follows. Section 2 describes the data and the conducted methodology. Section 3 reports and discusses the empirical estimation results. Section 4 concludes.

## 2. Data and methodology

*2.1.Data and descriptive statistics*

This study investigates the dynamic dependence between gold returns and an uncertainty composite index, conditional on different gold market states and various kinds of uncertainty levels. Our investigation is first based on the gold returns (Gr). We employ the first-differenced natural logarithm of the gold fixing price at 19:30 A.M. (London time), which is downloadable from the Federal Reserve Bank of St. Louis. As mentioned above, the composite indicator (CUCI) is constructed by summarizing six uncertainty factors. As an indicator for macroeconomic uncertainty (MAUCI), we utilize a measure developed by Jurado et al. (2015), based on a common factor derived from a panel incorporating the unforecastable component of a variety of monthly 132 macroeconomic time series. In addition, we employ an indicator of microeconomic uncertainty (MIUCI) developed by Bachmann et al. (2013), built on the forecast dispersion in the business climate. Besides, we utilize the index of economic policy uncertainty (EPUCI) provided by Baker et al. (2016) based on three main components: (a) the frequency of newspaper references to economic policy uncertainty by referring to articles containing the following keywords: "uncertainty", "economy", "congress", "deficit", "Federal Reserve", "legislation", "regulation" or "white house", (b) the tax provisions scheduled to expire , as well as (c) the disagreement among forecasters over the future government expenditure and inflation. We also account for a monetary policy uncertainty (MPUCI) recently proposed by Husted et al. (2016) able to detect the effectiveness of the Federal Reserve policy actions and their outcomes by searching for keywords related to monetary policy uncertainty in major newspapers. These keywords mainly include "uncertainty" or "uncertain," "monetary policy" or "interest rate", "policy rate" or "refinancing tender". Further, we consider a financial uncertainty index (FUCI) constructed by Ludvigson et al. (2015) based on the methodology of Jurado et al. (2015), by summarizing monthly 147 financial time series including bonds, stocks and commodity markets. Ultimately, a political uncertainty indicator (PUCI) developed by Azzimonti (2017) is accounted for. This index is developed by referring to the frequency of newspaper articles displaying disagreements among US politicians. Due to the availability of uncertainty proxies under study, the data cover the sample period from January 1999 to September 2015 (monthly frequency).[4] Although MAUCI, MIUCI, EPUCI, FUCI and PUCI are available for longer periods, MPUCI focuses on a restricted period from January 1999 to July 2017. Table 1 reports all the data used, their availability and their sources.

---

[4] We thank the Reviewer for the very careful review of our paper and for pointing out insightful remarks. A major revision of the paper has been carried out to take all of them into account. Specifically, we reestimate the dependence between gold returns and all the uncertainty proxies while fixing the frequencies and sample periods throughout the study. Because the alternative measures of uncertainty differ in terms of the data inputs of the uncertainty proxies, this study swaps to the index containing more data sources as soon as it is available. The present research considers similar frequencies and time periods for all the considered uncertainty proxies, i.e., monthly data for the period January 1999-September 2015.

**Table 1.** Data, definitions, availability and sources

| | Variables | Definition | Availability | Links of data sources |
|---|---|---|---|---|
| The dependent variable: | Gr | Gold returns | April 1968-May 2018 | https://fred.stlouisfed.org/series/GOLDAMGBD228NLBM |
| The independent variable: The uncertainty proxies | MAUCI | Macroeconomic uncertainty | July 1960-December 2017 | https://www.sydneyludvigson.com/data-and-appendixes/ |
| | MIUCI | Microeconomic uncertainty | October 1989-September 2015 | https://www.american.edu/cas/faculty/sheng.cfm |
| | EPUCI | Economic policy uncertainty | January 1985-November 2015 | http://www.policyuncertainty.com/ orhttps://fred.stlouisfed.org/series/USEPUINDXD |
| | MPUCI | Monetary policy uncertainty | January 1999-July 2017 | http://www.policyuncertainty.com/hrs_monetary.html |
| | FUCI | Financial uncertainty | July 1960-December 2017 | https://www.sydneyludvigson.com/data-and-appendixes/ |
| | PUCI | Political uncertainty | January 1981-March 2018 | https://www.philadelphiafed.org/research-and-data/real-time-center/partisan-conflict-index |

The data for the changes in Gr, MAUCI, MIUCI, EPUCI, MPUCI, FUCI and PUCI have been plotted in Fig A.1 (appendices). We observe that the period under study witnessed a heightened uncertainty surrounding economic, macroeconomic, microeconomic, monetary and financial policies and also unforeseen political events. For the same period, we note that gold exhibits a volatile behavior. More importantly, the graphical evidence shows that there are some periods where gold and uncertainty seem positively correlated. For instance, we note specific periods (in particular, the onset of global financial crisis) where a large increase in uncertainty is accompanied by a rise in Gr. This holds true for all the uncertainty indicators but with different magnitudes. Besides, March 2013 (the bailout of Cyprus's Banks) is marked by an increased financial and political uncertainty (FUCI and PUCI) associated with a rise in Gr. Such heterogeneity in the responses of gold return to uncertainty indicators set out the relevance of developing a composite index able to account for the core dynamics of uncertainty. Moreover, there are some periods where gold and uncertainty operate in opposite direction. This highlights the importance of analyzing this relationship under diverse scenarios.

The descriptive statistics are summarized in Table 2. We show that the average of the changes in all the time series is positive (with the exception of PUCI). The standard deviation values indicate that MPUCI and EPUCI fluctuate more largely than MAUCI, MIUCI, FUCI and PUCI. The non-normality as indicated by the Jarque-Bera test and the excess kurtosis (i.e., heavy tailed) motivates us to look at quantiles-based approaches. In addition, we evaluate whether the reaction of gold returns to the various uncertainty indicators is statistically different across distinct quantile levels. The results of Koenker and Xiao (2002) test, available for readers upon request, overwhelmingly reject the null hypothesis of slope equality for various quantiles of Gr. These findings reinforce the appropriateness of quantiles-based models over ordinary least squares (OLS) regression.

**Table 2.** Descriptive statistics

|  | Gr | MAUCI | MIUCI | EPUCI | MPUCI | FUCI | PUCI |
|---|---|---|---|---|---|---|---|
| Mean | 0.016111 | 0.006950 | 0.014810 | 0.021635 | 0.026599 | 0.020168 | -0.004518 |
| Median | 0.010141 | 0.004763 | 0.012721 | 0.018697 | 0.025766 | 0.022808 | -0.004798 |
| Std. Dev. | 0.038015 | 0.011823 | 0.022346 | 0.031061 | 0.039378 | 0.021458 | 0.024853 |
| Skewness | 0.330977 | 0.325140 | 0.195006 | 0.050090 | -0.086380 | -0.201764 | -0.156279 |
| Kurtosis | 3.216635 | 3.475480 | 3.626639 | 3.650033 | 3.675392 | 1.695175 | 3.377854 |
| Jarque-Bera | 40.32070 | 19.32866 | 24.77822 | 11.26361 | 11.49339 | 15.85592 | 20.43955 |
| p-value | 0.000183 | 0.011487 | 0.009700 | 0.069395 | 0.062891 | 0.000361 | 0.000883 |

Note: Std. Dev. symbolizes the Standard Deviation; p-value corresponds to the test of normality based on the Jarque-Bera test.

To better ascertain the usefulness of the quantile-based approaches, which are nonlinear, we conduct the BDS test of nonlinearity on the residuals recovered from this OLS model ($Gr_t = \beta^\theta UCI_t + \alpha^\theta Gr_{t-1} + \varepsilon_t^\theta$). The BDS test developed by Brock et al. (1996) was performed to test for the null hypothesis of independent and identical distribution to detect a non-random chaotic behavior. If the null hypothesis is rejected, this implies that the linear model is mis-specified. The BDS test results, reported in Table 3, reveal that all the variables under study are nonlinearly dependent which is one of the indications of chaotic behavior, and justifies the importance of nonlinear analysis of the dependence between gold return and the uncertainty indicators. The quantile-on-quantile regression accounts for possible nonlinearity, structural breaks and regime shifts as it shows distribution-to-distribution effects.

**Table 3.** The BDS test for detecting nonlinearity in the return series

| Embedding dimension (m) | Gr | MAUCI | MIUCI | EPUCI | MPUCI | FUCI | PUCI |
|---|---|---|---|---|---|---|---|
| 2 | 21.45*** | 23.45** | 18.87*** | 16.32** | 18.26** | 15.41** | 18.69** |
| 3 | 19.68** | 19.48** | 19.11*** | 18.21** | 19.43** | 17.05** | 17.13*** |
| 4 | 20.43** | 20.11*** | 18.34** | 19.13** | 21.22** | 19.86** | 18.72*** |
| 5 | 19.27** | 17.89* | 17.73*** | 21.46** | 20.17** | 18.65** | 18.09* |
| 2 | 18.78*** | 16.93** | 19.06*** | 19.67** | 19.24** | 17.21** | 17.67** |
| 3 | 18.19** | 17.74* | 20.43** | 21.33** | 22.58** | 18.18** | 19.14*** |
| 4 | 18.68*** | 18.12** | 19.96** | 19.16** | 24.15 | 21.04** | 20.23** |
| 5 | 18.22** | 18.04** | 19.02* | 20.48** | 19.42** | 21.68** | 20.91*** |

Notes: m denotes the embedding dimension of the BDS test. *, **, *** indicate a rejection of the null of residuals being i.i. d at 10%, 5 % and 1% levels of significance.

### 2.2. The quantile-on-quantile regression

The ability of gold to act as a hedge and a safe haven in uncertain times depends on how gold prices and uncertainty are related. To differentiate between the hedge and safe-haven properties, we determine the dependence between gold returns and changes in uncertainty levels in terms of average and joint extreme movements (Reboredo, 2013). Accurately, the gold is perceived as a hedge if it exhibits a positive link with uncertainty in normal states (i.e., when the uncertainty is at its middle level). It is seen as a safe haven if it is positively correlated with uncertainty when the uncertainty is higher.

This paper contributes to the literature on the same issue by using a quantile-on-quantile regression (QQR) approach developed by Sim and Zhou (2015).[5] The QQR allows capturing the dynamic connection between gold returns and uncertainty depending on different gold market conditions (i.e., bearish, normal or bullish) and the nuances of uncertainty(i.e., low, middle or high).The evidence regarding the tail-dependence between the quantile of gold returns and the quantile of changes in uncertainty is prominent for investors hedging an uncertain exposure. The QQR is a generalization of the standard quantile regression (QR) approach, which is able to examine how the quantiles of a variable affect the conditional quantiles of another variable (Sim and Zhou, 2015). The QQR helps to assess the entire dependence structure of gold returns and uncertainty by using their return information. Specifically, by using the QQR, we are able to model the quantile of gold returns (Gr) as a function of the quantile of changes in uncertainty (UCI), so that the link between these time series could vary at each point of their respective distributions. More details about the differences between OLS, QR and QQR are provided in Fig.1.

---

[5]The R package quantreg.nonpar was utilized to implement the quantile-on-quantile regression, which is nonlinear. The R code used in this study is promptly available for readers upon request.

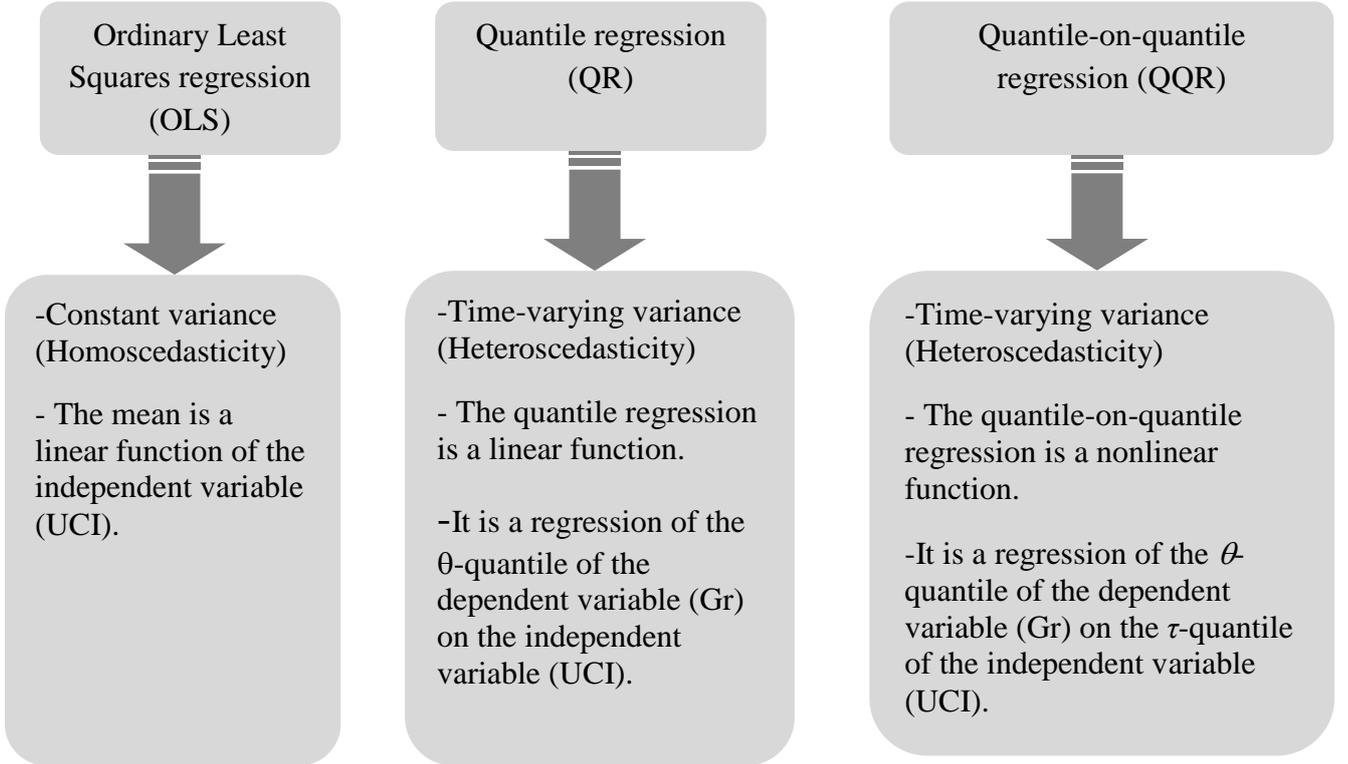

**Fig. 1.** OLS vs. quantile regression vs. quantile-on-quantile regression

Let superscript $\theta$ denotes the quantile of gold returns (Gr). We first postulate a model for the $\theta$-quantile of Gr as a function of the uncertainty index (UCI) and the lagged Gr.

$$Gr_t = \beta^\theta UCI_t + \alpha^\theta Gr_{t-1} + \varepsilon_t^\theta \tag{1}$$

where $\varepsilon_t^\theta$ is an error term that has a zero $\theta$-quantile. We allow the relationship function $\beta^\theta$(.) to be unknown, as we do not have prior information of how Gr and UCI are related. To examine the dependence structure between the quantile of Gr and the quantile of UCI, denoted by UCI$^\tau$, we linearize the function $\beta^\theta$(.) by considering the first-order Taylor expansion of $\beta^\theta$(.) around UCI$^\tau$. We have:

$$\beta^\theta(UCI_t) \approx \beta^\theta(UCI^\tau) + \beta^{\theta'}(UCI^\tau)(UCI_t - UCI^\tau) \tag{2}$$

Sim and Zhou (2015) define $\beta^\theta(UCI^\tau)$ and $\beta^{\theta'}(UCI^\tau)$, respectively, as $\beta_0(\theta,\tau)$ and $\beta_1(\theta,\tau)$. The Eq (2) can be, thereafter, expressed as follows:

$$\beta^\theta(UCI_t) \approx \beta_0(\theta,\tau) + \beta_1(\theta,\tau)(UCI_t - UCI^\tau) \tag{3}$$

The following step consists of substituting the Eq (3) into the Eq (1) to get:

$$Gr_t = \beta_0(\theta,\tau) + \alpha(\theta)Gr_{t-1} + \beta_1(\theta,\tau)(UCI_t - UCI^\tau) + \varepsilon_t^\theta \tag{4}$$

where $\alpha^\theta \equiv \alpha(\theta)$, with Eq (4) carried to the $\theta$-conditional quantile of *Gr*. Unlike the standard quantile regression, the expression $\beta_0(\theta,\tau) + \beta_1(\theta,\tau)(UCI_t - UCI^\tau)$ considers the linkage between the $\theta$-quantile of Gr and the $\tau$-quantile of UCI, given that $\beta_0$ and $\beta_1$ are doubly indexed in $\theta$ and $\tau$. In short, the QQR allows for detecting more completely the dependence between the Gr and UCI through the dependence between their respective distributions.

While attempting to estimate Eq (4), we solve

$$\min_{\beta_0 \beta_1} \sum_{i=1}^n \rho_\theta \left[ Gr_t - \beta_0 - \beta_1(UCI_t - UCI^\tau) - \alpha(\theta)Gr_{t-1} \right] K\left( \frac{F_n(UCI_t) - \tau}{h} \right) \tag{5}$$

where $\rho_\theta$ corresponds to the absolute value function that gives the θ-conditional quantile of Gr as a solution. We then conduct a Gaussian kernel $K(.)$ to weight observations according to a normal probability distribution based on the bandwidth h=0.05. The weights seem reversely linked to the distance of $UCI_t$ from $UCI^\tau$:

$$F_n(UCI_t) = \frac{1}{n}\sum_{k=1}^n I(UCI_k < UCI_t) \tag{6}$$

from $\tau$, where $\tau$ is the value of the distribution function that refers to $UCI^\tau$.

We pursue the same exercise to assess the dependence between gold returns and the six aforementioned uncertainty proxies (i.e., MAUCI, MIUCI, EPUCI, MPUCI, FUCI and PUCI) as well as the developed composite uncertainty indicator.

## 3. The empirical results

### *3.1. The quantile-on quantile results*

This study focuses on the direct relationship between gold returns and uncertainty conditional on the different gold market circumstances and the nuances of uncertainty levels. Specifically, we consider that quantiles reflect how bearish, normal or bullish gold market is and whether the uncertainty index is low, normal or high. To investigate the dynamic dependencies between gold returns and changes in uncertainty during the bear (bull) states, the linkages between the 10th, 20th, 30th and 40th (60th, 70th, 80th and 90th) return quantiles are considered. The return dependencies during the normal state are determined through the centrally located quantiles (50th return quantile).

Based on the quantile-on-quantile regression approach expressed in Eq (4), the entire dependence between the quantile of gold return (indexed by *θ*) and the quantile of uncertainty indicators (indexed by *τ*) can be synthesized by two main parameters: $\beta_0(\theta, \tau)$ and $\beta_1(\theta, \tau)$, the intercept term and the slope coefficient, respectively. Being function of *θ* and *τ*, both parameters vary depending to the different gold market states and the nuances of uncertainty levels. The left side of Fig 2 plots the surface of the intercept term. Unlike the intercept term of the standard quantile regression that are insensitive to uncertain quantiles, the intercept derived from QQR depends to both the quantiles of Gr in the Y-axis and the quantiles of UCI in the X-axis. We note that the intercept terms are generally positive and greater at the bottom quantiles for MAUCI, EPUCI, MPUCI, FUCI and PUCI (i.e., when the uncertainty is low) and at upper quantiles for MIUCI (i.e., when the uncertainty is high), and under various gold market circumstances. For example, for the relationship between Gr and MAUCI, the intercept parameter seems positive and strong under bull gold market states (*θ*=0.7, 0.8, 0.9) and when the uncertainty is low (*τ*=0.2, 0.3, 0.4). It is, however, negative when the gold market is bearish (*θ*=0.2) and the uncertainty is low (*τ*=0.3, 0.4). When looking at the dependence of Gr and MIUCI, the intercept term is positive and strong under normal and bull gold market conditions ($\theta$=0.5, 0.6, 0.7) and when the uncertainty is high ($\tau$=0.7, 0.8, 0.9). It is also positive when the gold market is improving ($\theta$=0.6, 0.7) and the uncertainty is declining ($\tau$=0.3). Nevertheless, the intercept is likely to be negative when the uncertainty attains its highest level ($\tau$=0.9) and during bear gold market conditions ($\theta$=0.3).

The right side of Fig 2 plots the gold return's response to different uncertainty proxies. Interestingly, as documented by the previous studies on the same issue, we show that the different uncertainty measures display heterogenuous impacts on gold returns (for example, Blose, 2010; Jones and Sackley, 2016; Balcilar et al. 2016; Beckmann et al. 2017).Fig 2 (a) reports the estimate of the gold return's reaction to the macroeconomic uncertainty. A pronounced positive response is observed when the macroeconomic uncertainty is high ($\tau$=0.6, 0.7, 0.8) and under bear and normal gold market regimes ($\theta$=0.2, 0.3, 0.4, 0.5).However, a negative and strong relationship is observed when the uncertainty is around the normal ($\tau$=0.5) and the bull ($\theta$=0.6, 0.7, 0.8) gold market states. When considering the microeconomic uncertainty (Fig 2(b)), we show a positive and great response of Gr when the MIUCI attains its highest level ($\tau$=0.9) and when the gold market is bearish ($\theta$=0.4) or normal ($\theta$=0.5). This connection remains positive when the uncertainty is high ($\tau$=0.7) and under normal ($\theta$=0.5) and mildly bull ($\theta$=0.6, 0.7) gold market circumstances. This linkage is likely to be negative and pronounced when the uncertainty is mildly high ($\tau$=0.6, 0.7) and under bull ($\theta$=0.6, 0.7, 0.8, 0.9) gold market states. Moreover, the response of gold returns to the economic policy uncertainty reported in Fig 2(c) seems positive and substantial when the EPUCI is high ($\tau$=0.6, 0.7, 0.8, 0.9) and under bear gold market scenario ($\theta$=0.2, 0.3). A negative link is, nevertheless, seen when the uncertainty is low or middle ($\tau$=0.2, 0.3, 0.4, 0.5) and under normal and bull gold market circumstances ($\theta$=0.5, 0.6, 0.7, 0.8). Further, we find that gold returns react positively and widely to the monetary policy uncertainty (Fig 2(d)) when the MPUCI is highest ($\tau$=0.9) and under various gold market circumstances ($\theta$=0.4, 0.5, 0.6, 0.7). A negative Gr-MPUCI correlation is shown when the uncertainty is high ($\tau$=0.8) and under bull gold market state ($\theta$=0.9). If we consider the financial uncertainty (Fig 2(e)), we note a positive and pronounced Gr reaction under heightened uncertainty period ($\tau$=0.7, 0.8) and when the gold market is normal ($\theta$=0.5) or bullish ($\theta$=0.7, 0.8, 0.9). This dependence seems also positive but with less extent when the uncertainty is low ($\tau$=0.2) and under bull gold market regimes ($\theta$=0.7, 0.8, 0.9). Nevertheless, this relationship appears negative whatever the uncertainty level (i.e., low, normal or high) and when the gold market is declining ($\theta$=0.1, 0.2, 0.3). By focusing on the effect of political uncertainty on gold price dynamics (Fig 2(f)), we clearly show a positive and strong Gr response (a) when the uncertainty is highest ($\tau$=0.9) and the gold market is bearish ($\theta$=0.2, 0.3, 0.4), (b) when the uncertainty is normal or mildly high ($\tau$=0.5, 0.6) and under distinct gold market conditions (bear: $\theta$=0.2, 0.3, 0.4; normal: $\theta$=0.5; bull: $\theta$=0.6, 0.7). However, a negative correlation is revealed when the uncertainty is great ($\tau$=0.8) and the gold market is functioning around its normal condition ($\theta$=0.5).

Fig 2 (a).Gr and MAUCI

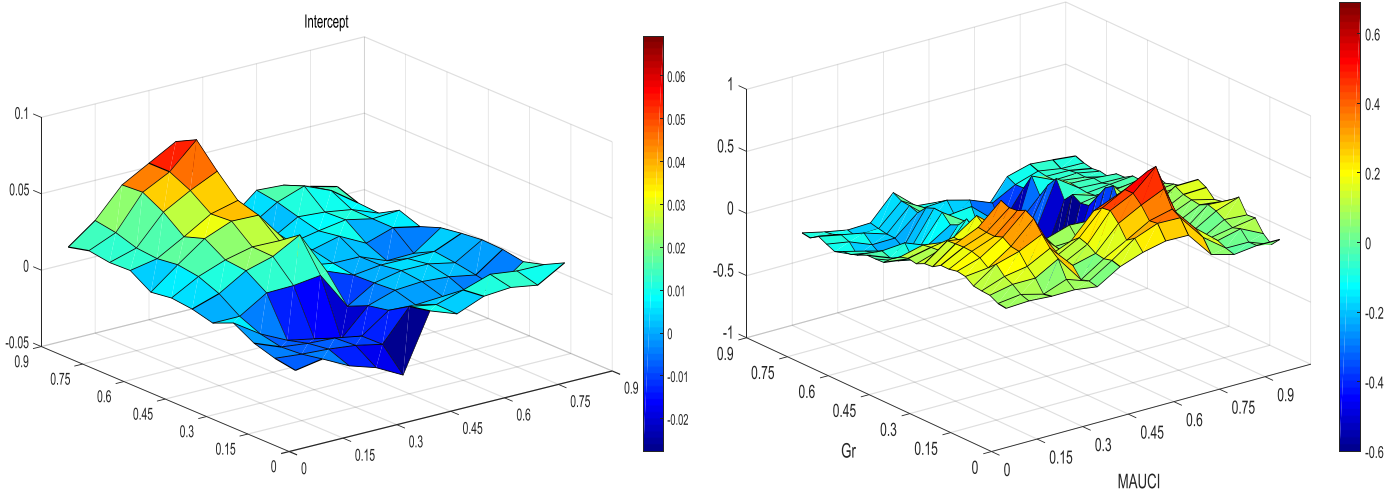

Fig 2 (b).Gr and MIUCI

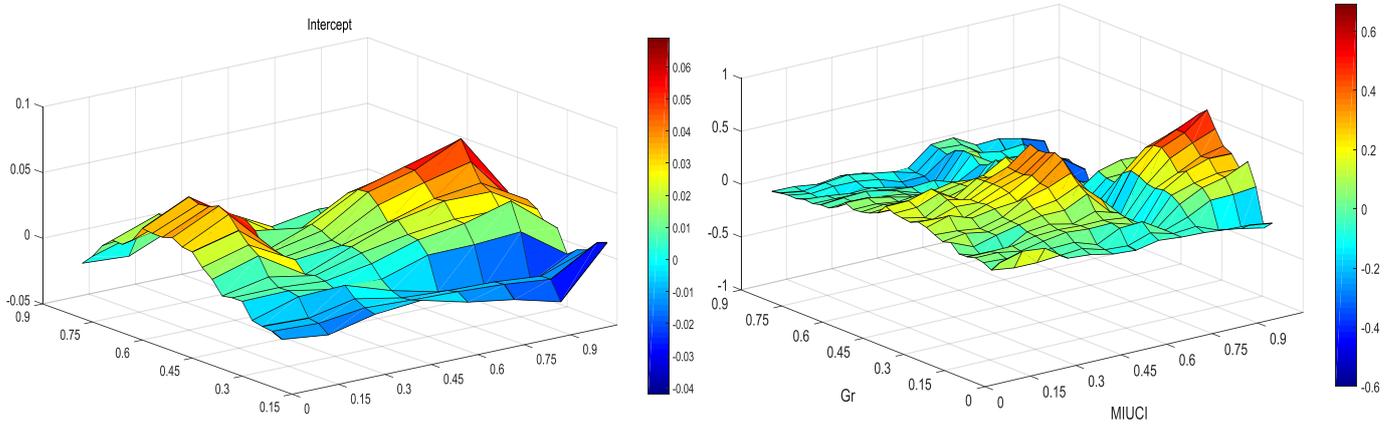

Fig 2(c).Gr and EPUCI

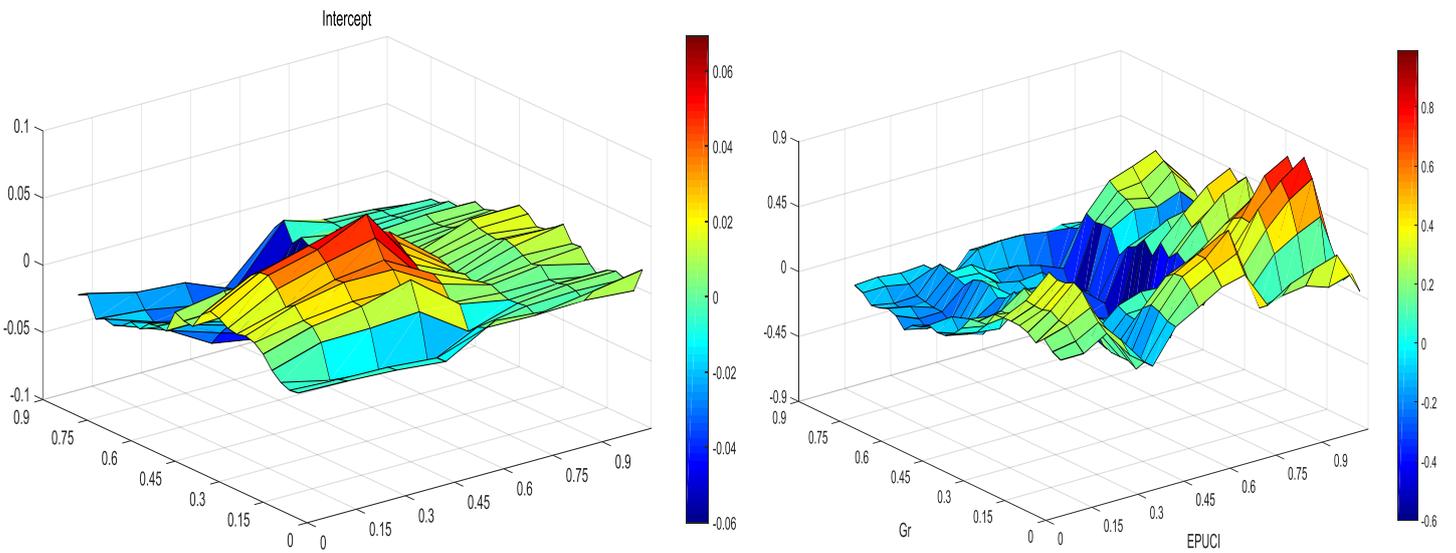

Fig 2 (d).Gr and MPUCI

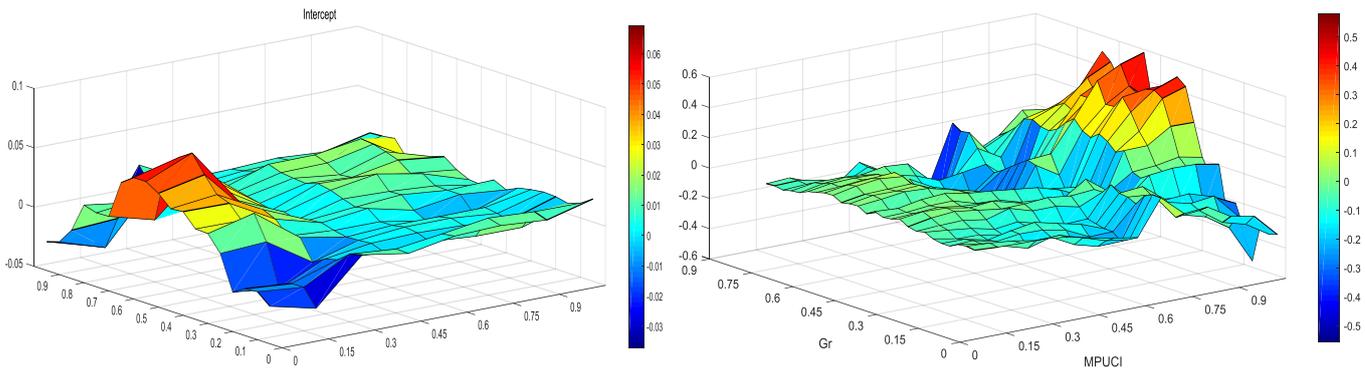

Fig 2 (e).Gr and FUCI

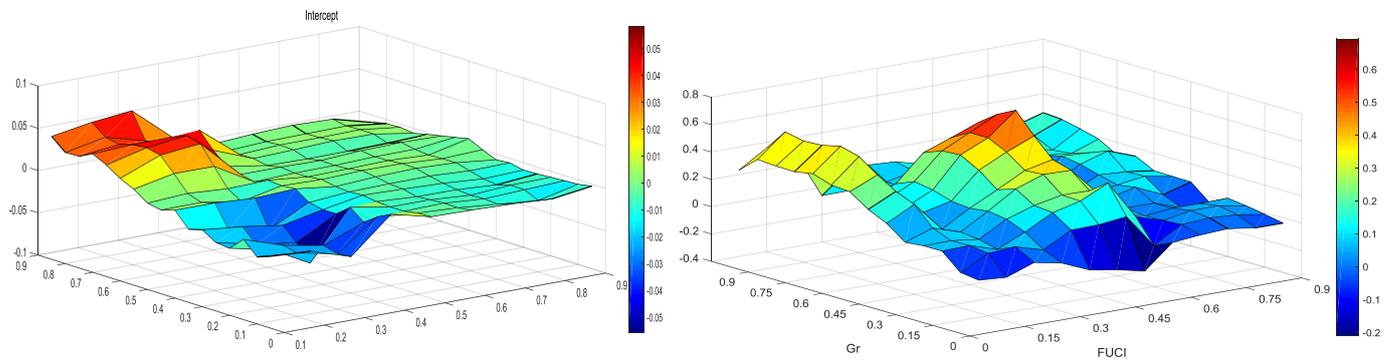

Fig 2 (f).Gr and PUCI

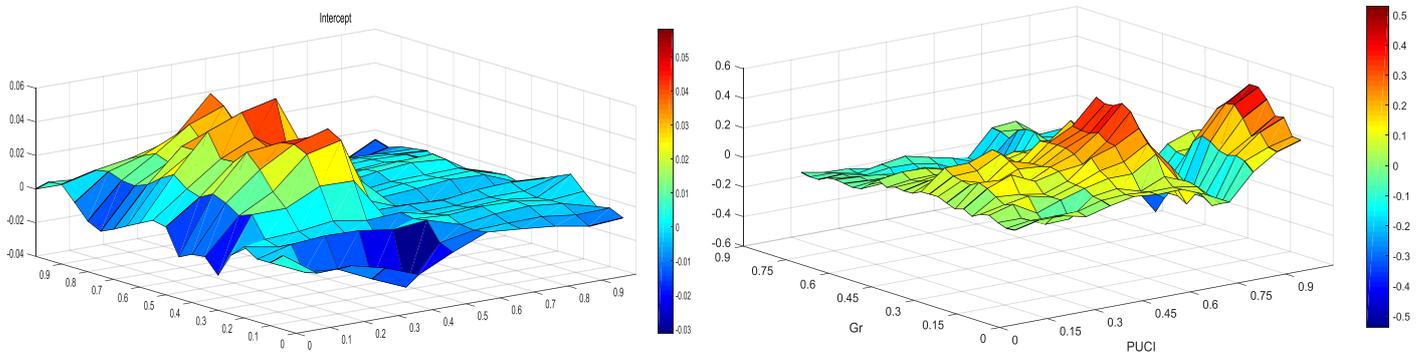

**Fig. 2.** Gold returns and uncertainty indicators: The estimated parameters of the quantile-on-quantile regression

Note: This figure shows the estimated parameters of Eq (4). The left-side of the figure plots the change in the intercept term $\beta_0(\theta, \tau)$ in the z-axis against the θ-quantile of gold return and the τ-quantile of uncertainty in the x-y axes. The right side of the figure depicts the estimates of the slope coefficient, $\beta_1(\theta, \tau)$, which is placed on the z-axis against the quantiles of the Gr (θ) on the y-axis and the quantiles of UCI (τ) on the x-axis. For both the intercept parameter and the slope coefficient, the red (yellow) color corresponds to positive and strong (weak) values of the intercept/ the slope coefficient, while the dark (light) blue color corresponds to negative and pronounced (moderate) values of the intercept/ the slope coefficient. A light green color corresponds to very modest or negligible values of the intercept/the slope coefficient.

Overall, our results indicate that gold serves as a safe haven against all the types of uncertainty, but as a hedge solely against the political uncertainty. These properties depend also to the gold market conditions. The above mixed findings which can be traced back to the various definitions and determinants of these uncertainty indicators. This has motivated us to construct an uncertainty composite indicator (CUCI) by summarizing the aforementioned uncertainty indexes. This allows us to investigate the core effects of uncertainty on gold price dynamics conditional on the different gold market circumstances and the various levels of uncertainty.

### 3.2. Construction of an uncertainty composite indicator

Having identifying heterogenuous reactions of gold returns to the different uncertainty proxies used throughout this study, we try in the following to determine their common driving forces which we interpret as the composite uncertainty indicator. To this end, we first carry out a factor model written as follows:

$$X_t = \lambda f_t + \zeta_t \tag{7}$$

$$f_t = \psi(L) f_{t-1} + \upsilon_t \tag{8}$$

Each $X_t$ corresponds to the sum of two unobservable components, a common factor detecting the core uncertainty effects ($x_t = \lambda f_t$) and an idiosyncratic component ($\zeta_t$) capturing specific shocks or measurement errors. Note that the common factor and the idiosyncratic component are uncorrelated ; The data vector ($X_t = \log(\sigma_{1,t}),..., \log(\sigma_{n,t})$) of dimension n × 1 incorparates the standardized individual uncertainty measures, i.e., economic, macroeconomic, microeconomic, monetary policy, financial and political uncertainty factors. For the factor model, we employ the logarithmic square roots of the variances ($\log(\sigma_{1,t}),..., \log(\sigma_{n,t})$) to enable the factors to have negative values ; $f_t = (f_{1t}, ..., f_{nt})$ is an n-dimensional vector of common factors influencing all the uncertainty indicators ; $\lambda = (\lambda_{1t}, ...,$ $\lambda_{nt}$) is an n-dimentional vector of factor loadings where each of the element in $\lambda$ potentially reflects the impact of the common factor to uncertainty indexes under study.

Thereafter, we let $f_t$ in Eq (8) to pursue a vector autoregressive process. The lag polynomial $\psi(L) = \psi_1 + \ldots + \psi_p L^{p-1}$ [6] is of dimension r × r. The corresponding innovations are expressed by the r × 1 dimensional vector $v_t$, and can be disentangled into $v_t = R\mu_t$. The r-dimensional vector $\mu_t$ incorporates orthogonal white noise shocks and $R$ is an r × r matrix. We should stress at this stage that the factor innovations and idiosyncratic components are presumed to be independent at all lags (*L*). We suppose that the number of fundamental shocks $\mu_t$ is similar to the that of the common factors *r*.

Ultimately, we estimate the Eqs (7) and (8) by means of the Quasi Maximum Likelihood (QML) procedure developed by Doz et al. (2012). Methodologically, The QML is based on an EM-algorithm combined with a Kalman smoother. More accurately, Doz et al. (2012) computed the EM algorithm with an estimate of $\hat{f}_t$ derived from the principal components corresponding to the *r* highest eigenvalues of the covariance matrix of $X_t$. We, thereafter, iterate between two main steps : M-step and E-step (see Fig. 3).[7]

---

[6] In general, we define the lag operator by : $Lx_t = x_{t-1}$ and $L^k x_t = x_{t-k}$. The lag polynomials is considered as polynomials in the lag operator. Let $\psi(L) = \psi_1 + \ldots + \psi_p L^{p-1}$ be the lag polynomial. As an operator, it is defined by $\psi(L)x_t = \psi_1 x_{t-1} + \psi_2 . x_{t-2} + \ldots + \psi_p x_{t-p}$.

[7] We build a set of Matlab routines to run the dynamic factor model (DFM). The source code is available on this link: http://www.barigozzi.eu/Codes.html.

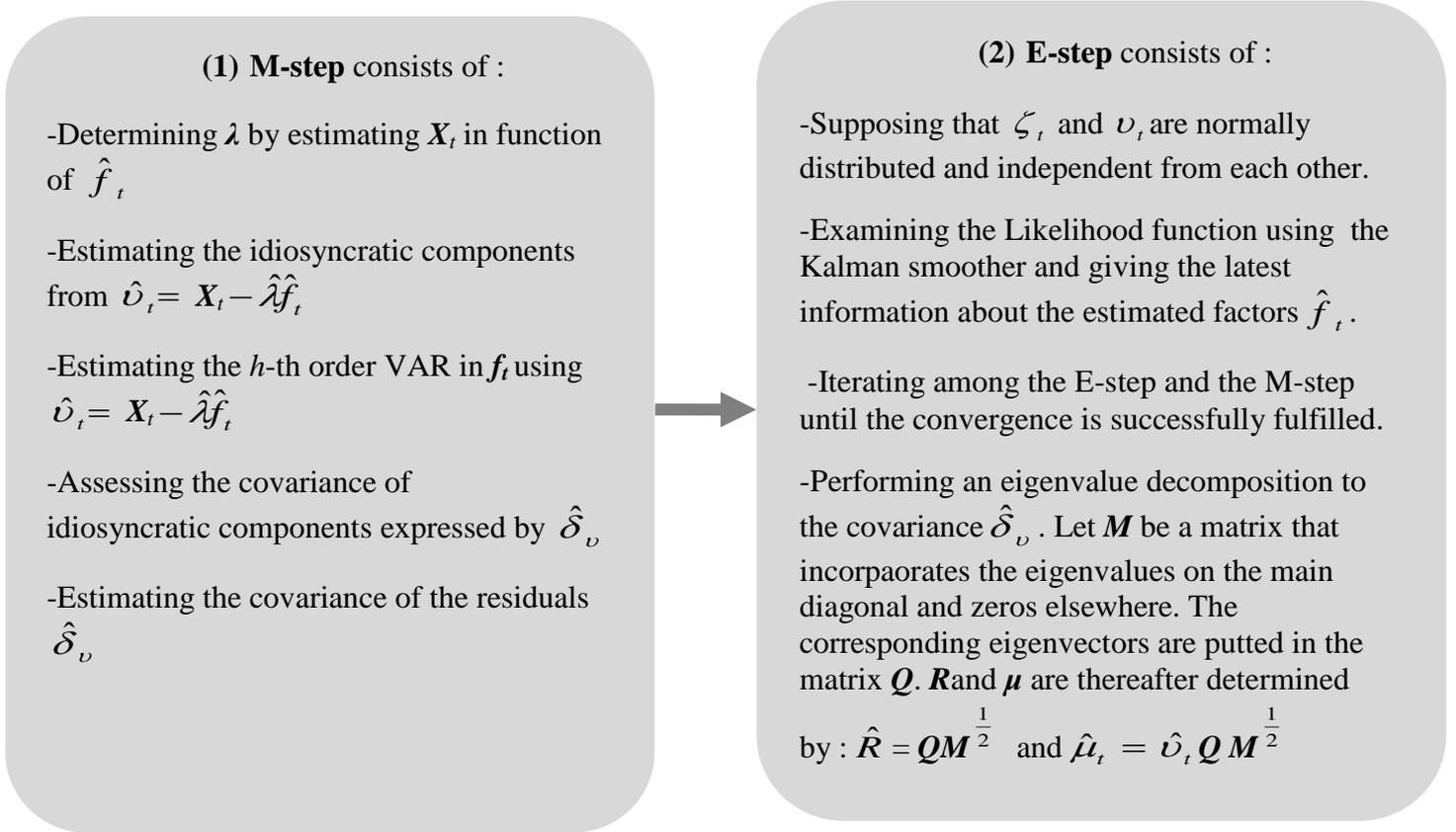

**Fig. 3.** An estimation of the dynamic factor model

The DFM has at least three main advantages. Firstly, it permits to synthesize the information incorporated in a wide dataset in a handful of common factors (Hezel and Malte, 2013). The latter are then utilized as measures of the uncertainty composite indicator. This will help to determine the full uncertainty's effects on gold returns, which are not specific to one measure of uncertainty. Secondly, unlike the principal component analysis, the DFM properly detects the variability of components (Lim and Nguyen, 2015). Thirdly, the maximum likelihood approach is more appropriate for small samples.

*3.3. Measuring the response of gold returns to the uncertainty composite index*

Table 4 displays the correlation matrix among the different categories of uncertainty indexes under study. We show that all the uncertainty indicators used throughout this paper (i.e., MAUCI, MIUCI, EPUCI, MPUCI, FUCI and PUCI) exhibit positive and strong first-order autocorrelation, underscoring that macroeconomic, microeconomic, economic policy, monetary policy, financial and political uncertainty is persistent and that those indicators are positively related. The fact that these uncertainty proxies tend to move together, underscores that there exists a common uncertainty component to all the indicators. We thus try to determine this common uncertainty component by developing an uncertainty composite index (*CUIC*) using the dynamic factor approach. Because the alternative measures of uncertainty differ in terms of the data inputs of the uncertainty proxies and the methodologies conducted for constructing the indicators, this study swaps to the index containing more data sources as soon as it is available. In particular, to construct the composite uncertainty indicator, we use monthly data covering the period from January 1999 to September 2015.

**Table 4.** The correlation matrix between uncertainty proxies

|       | MAUCI | MIUCI   | EPUCI   | MPUCI   | FUCI    | PUCI    |
|-------|-------|---------|---------|---------|---------|---------|
| MAUCI | 1     | 0.97993 | 0.92974 | 0.86110 | 0.69737 | 0.19012 |
| MIUCI |       | 1       | 0.97952 | 0.93397 | 0.53592 | 0.23007 |
| EPUCI |       |         | 1       | 0.98279 | 0.47910 | 0.27314 |
| MPUCI |       |         |         | 1       | 0.41725 | 0.30773 |
| FUCI  |       |         |         |         | 1       | 0.23083 |
| PUCI  |       |         |         |         |         | 1       |
| $\rho(1)$ | 0.73 | 0.88 | 0.76 | 0.63 | 0.71 | 0.54 |

Note : $\rho(1)$ is the first-order autocorrelation.

Table 5 reports the descriptive statistics for the uncertainty composite index. Based on the dynamic factor model, we find that all the economic, macroeconomic, microeconomic, monetary policy, financial and political uncertainties are responsible for the common uncertainty dynamics (strong correlation between the different uncertainty measures and the composite index, with less extent for PUCI). The common variation in uncertainty is strongly explained by MPUCI followed by MAUCI. This result suggests that for effective policy response to increased uncertainty, it is crucial to account for various sources of uncertainty.

**Table 5.** The descriptive statistics for the composite uncertainty indicator

|                   | CUCI    |
|-------------------|---------|
| *Basic statistics*|         |
| Mean              | 0.01156 |
| Std.Dev.          | 0.06852 |
| Skewness          | 0.98632 |
| Kurtosis          | 3.18725 |
| *Correlations*    |         |
| MAUCI             | 0.78    |
| MIUCI             | 0.68    |
| EPUCI             | 0.61    |
| MPUCI             | 0.81    |
| FUCI              | 0.54    |
| PUCI              | 0.19    |
| $\rho(1)$         | 0.94    |

Note: Std. Dev. symbolizes the Standard Deviation; $\rho(1)$ is the first-order autocorrelation.

Fig. 4 describes the evolution of the composite uncertainty index developed based on the dynamic factor model. We clearly notice that the synthetic composite index coincides with the well-known heightened uncertainty periods including the terrorist attacks of September 2001, the onset of the global financial collapse (2008-2009), the Greek debt crisis, the bailout of Cyprus's Banks (March 2013) and the China's economic slowdown (since the first quarter of 2015). However, when comparing the levels of CUCI with the different uncertainty proxies used throughout our study, we notice some dissimilarities (see Fig A1), such as during the Chinese economic downturn; While CUCI increases remarkably, we observe a drop in PUCI. These differences can be attributed to the fact that the uncertainty surrounding the Chinese crisis is specific to one source of uncertainty and is not common to the six sources of uncertainty under consideration; hence the usefulness of the composite index.

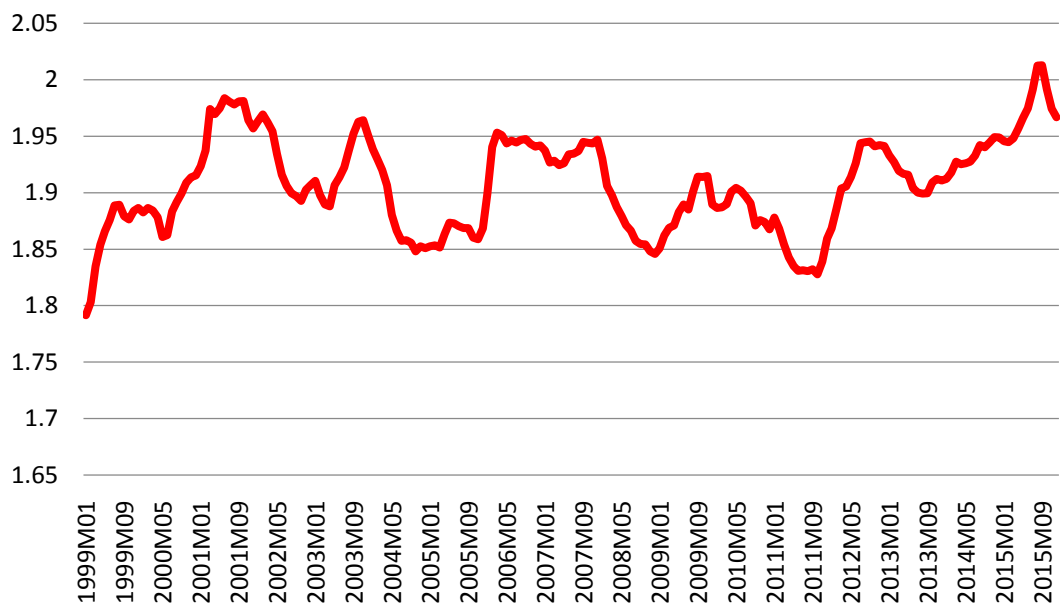

**Fig. 4.** The evolution of the uncertainty composite index

Using this developed synthetic index, we can assess more completely the role of uncertainty on gold price changes. Fig 5 shows the estimated parameters of the regression of the θ-quantile of gold return on the τ-quantile of the composite uncertainty index. The left-side of the figure plots the change in the intercept parameter. We note a positive (negative) and strong intercept term when the gold market is bullish (bearish) and under various uncertainty episodes.

The right side of this figure describes the dependence structure between gold returns and the uncertainty composite indicator for distinct combinations of $\theta$ and $\tau$ quantiles for Gr and CUCI, respectively. We show that there is a positive and strong relationship between gold returns and the uncertainty composite index when the uncertainty reaches its highest level ($\tau=0.9$) and when the gold market is bear or normal ($\theta=0.4, 0.5$). A moderately positive Gr response to CUCI is seen (a) when the uncertainty is high ($\tau=0.9$) and under bullish gold market regime ($\theta=0.6, 0.7, 0.8, 0.9$), (b) when the uncertainty level is around the average ($\tau=0.5$) or mildly high ($\tau=0.6, 0.7$) and the gold market is bear or normal ($\theta=0.3, 0.4, 0.5$), and (c) when the uncertainty is low ($\tau=0.3, 0.4$) and under different gold market states (bear:

$\theta$=0.4; normal:$\theta$=0.5; bull: $\theta$=0.6). We observe, however, a negative and wider reponse of Gr to CUCI when the uncertainty level is low or middle ($\tau$=0.4, 0.6) and under normal ($\theta$=0.5) and bull ($\theta$=0.6, 0.7, 0.8, 0.9) gold market conditions. It must be stressed here that the dependence between the uncertainty composite indicator and gold returns is more intense than the correlations between the individual uncertainty indexes and Gr, highlighting the relevance of the composite index. The positive and strong effect of uncertainty on gold returns when uncertainty attains its highest level is not surprising since an increase in uncertainty due to an unexpected event that caused significant macroeconomic, microeconomic, economic, monetary policy, financial or political instability might materialize in sharp changes in the price of a safe haven asset. This could occur since market participants (investors or traders) respond to the great uncertainty by rebalancing their investments toward the safe asset, or because those who hold such an asset are less willing to sell it (Caballero and Krishnamurthy, 2008; Piffer and Podstawski, 2017). But our findings contribute to the existing literature on the issue by showing that this relationship is positive (negative) when the uncertainty is highest or middle (low) and under different the gold market states. While relatively unambiguous, our results suggest that the gold's role as a hedge and safe haven cannot necessarily hold at all times. We can attribute this outcome to the mood of gold markets, the fundamentals of supply and demand in the gold market, the behavior of other assets as well as the effectiveness of conventional and unconventional monetary policy measures.

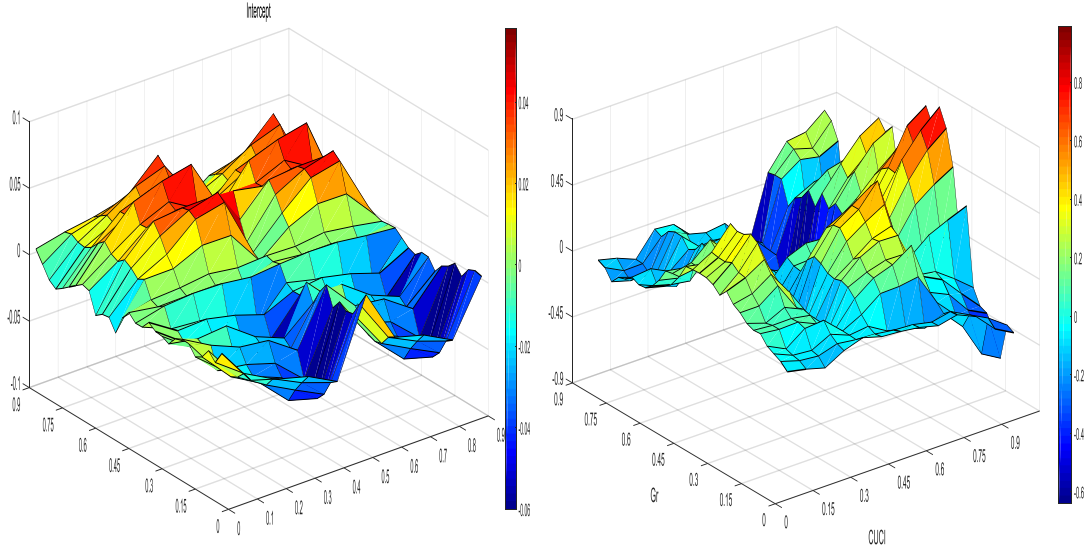

**Fig. 5.** Gold returns and composite index: The estimated parameters of the quantile-on-quantile regression

Note: This figure shows the estimated parameters of Eq (4). The left-side of the figure plots the change in the intercept term $\beta_0(\theta, \tau)$ in the z-axis against the θ-quantile of gold return and the τ-quantile of uncertainty in the x-y axes. The right side of the figure depicts the estimates of the slope coefficient, $\beta_1(\theta, \tau)$, which is placed on the z-axis against the quantiles of the Gr (θ) on the y-axis and the quantiles of CUCI (τ) on the x-axis. For both the intercept parameter and the slope coefficient, the red (yellow) color corresponds to positive and strong (weak) values of the intercept/ the slope coefficient, while the dark (light) blue color corresponds to negative and pronounced (moderate) values of the intercept/ the slope coefficient. A light green color corresponds to very modest or negligible values of the intercept/ the slope coefficient.

*3.4. The quantile-on-quantile regression vs. the standard quantile regression*

The QQR consists of decomposing the standard quantile regression (QR) estimates so that they are specific parameters for the several quantiles of the independent time series. Indeed, the QR cannot appropriately capture the entire dependence between gold returns and uncertainty. Although the QR seems able to estimate the distinct responses of gold returns to uncertainty at various points of the conditional distribution of gold, it overlooks that the level of uncertainty might also exert a significant influence on the gold's hedge and safe haven benefits. To check the effectiveness of the QQR approach compared to the standard QR, we attempt in the following to compare the QR estimates with the *τ*-averaged QQR parameters. But before starting this assessment, it must be pointed out that the QQR regresses the *θ*-quantile of gold returns on the *τ*-quantile of uncertainty (double indexing, i.e., *θ* and *τ*), whereas the QR regresses the $\theta$-quantile of gold returns on the changes in uncertainty (solely indexed by $\theta$). This implies that by carring out a QQR, we can provide better insights and more information on the response of gold returns to uncertainty.

To construct now the parameters from the QQR model that are indexed by $\theta$, the estimated QQR parameters were displayed by averaging along $\tau$. Consequently, the impact of uncertainty indexes on the distribution of gold returns is denoted by $\hat{\gamma}_1(\theta)$:

$$\hat{\gamma}_1(\theta) = \frac{1}{s} \sum_{\tau} \hat{\beta}_1(\theta, \tau) \tag{9}$$

where $s=48$ is the number of quantiles $\tau = \{0.02, 0.04, ..., 0.98\}$.

Fig. 6depicts the trajectory of the quantile regression and averaged QQR estimates of the slope coefficient that measures the effect of the uncertainty composite index on gold returns. We notice that whatever the quantile level (i.e., bottom: $\tau$ =0.1, 0.2, 0.3, 0.4, middle: $\tau$ =0.5 or upper: $\tau$=0.6, 0.7, 0.8, 0.9), the averaged QQR estimates of the slope coefficient are likely to be similar to the quantile regression estimates. This graphical illustration provides a simple validation of the QQR methodology by revealing that the main features of the quantile regression model can be recovered by summarizing the detailed information incorporated in the QQR estimates. Nevertheless, the QQR may offer more insights about the response of gold returns to uncertainty than QR as the latter does not controls for the possibility that whether the uncertainty level is low, middle or high may also affect on the way uncertainty indexes and gold returns are significantly interlinked.

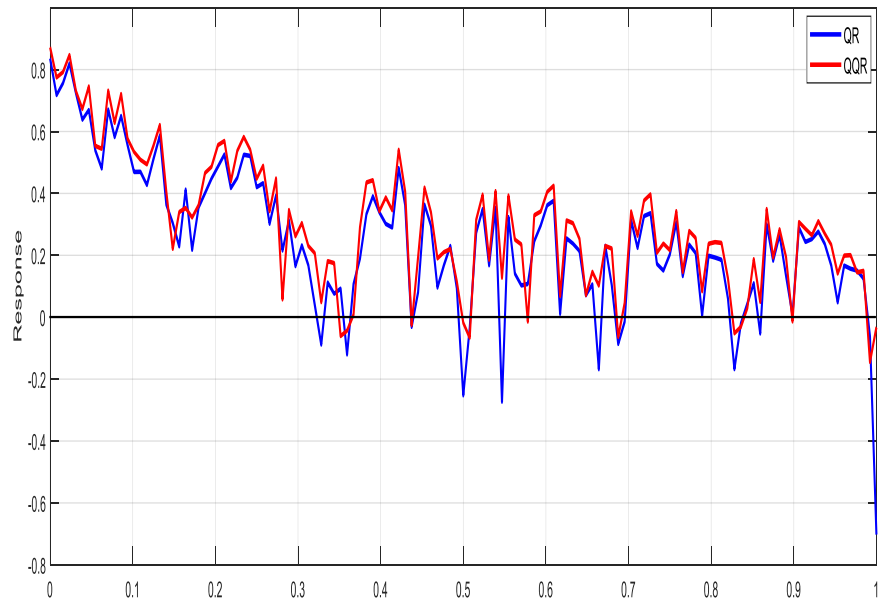

**Fig. 6.** The dependence structure between gold returns and the composite uncertainty index: QR vs. QQR

## 4. Conclusions

This paper extends the common focus on the correlation between gold and assets such as stocks or bonds with an analysis of the response of gold returns to uncertainty. Although some recent studies attempt to explore the link between gold price dynamics and uncertainty, no research to date has examined the core effects of uncertainty on the path of the gold price while paying specific attention to tail dependence. We use a quantile-on quantile regression to investigate the reaction of gold returns to various uncertainty proxies under different gold market conditions and distinct uncertainty episodes. Having found heterogeneous gold reactions to the different uncertainty indicators, we propose a composite indicator that synthesizes the economic, macroeconomic, microeconomic, monetary policy, financial and political effects of uncertainty within one measure of uncertainty.

Our results indicate that there is a positive and strong dependence between gold returns and the composite uncertainty index in high uncertainty episodes. It is not surprising that gold provides great protection against heightened uncertainty. This yellow metal has a long been perceived as a hedge and effective safe haven used to store value, and has never gone to zero in recorded history. But the novelty of this study relies on deducing that (i) the ability of gold to serve as a hedge or a safe haven in uncertain episodes depends on the varying gold market circumstances, the nuances of uncertainty and the uncertainty measures used, and (ii) the relationship between gold and uncertainty is nonlinear, probably due to the mood of gold traders, the interplay of supply and demand in the gold market, the price fluctuations of other assets and the unconventional and conventional monetary policies.

Our findings are very timely and useful for both individual and institutional investors as the global financial markets continue to be persistently rocked by unpredictable and extremely destabilizing events. In today's uncertain environment, measuring the price dynamics of gold, which has a long history of being a reliable store of wealth, under different scenarios becomes fundamental in designing sensible risk management strategies. When situations of increased uncertainty arise, an effective defense is to be well informed. Throughout this study, we detail the risk facing market participants conditional on different sources of uncertainty, and provide them with information on various circumstances, especially how to deal with worst case conditions.

The present research only explores the relationship between gold price and uncertainty, but can be extended to other major precious metals (in particular, silver, palladium and platinum). Despite extensive research and interest in the role of precious metals to protect against risk and heightened uncertainty, the empirical literature on hedging capabilities of precious metals is biased towards the assessment of gold's properties.

Beckmann, J., Berger, T., and Czudaj, R., 2017. Gold Price Dynamics and the Role of Uncertainty. Chemnitz Economic Papers, No. 006 (May).

Bohl, M.T., Döpke, J., and Pierdzioch, C., 2009. Real-Time Forecasting and Political Stock Market Anomalies: Evidence for the United States. Financial Review 43, 323–335.

Bouoiyour, J., Selmi, R., and Wohar, M., 2017. Are Bitcoin and gold Prices Correlated? New Evidence from a Quantile-on-Quantile Regression Model. Working paper, CATT, University of Pau, France (November).

Brock, W. A., Scheinkman, J. A., Dechert, W. D., LeBaron, B., 1996. A test for independence based on the correlation dimension. Econometric Reviews, 15(3), 197-235.

Caballero R.J., and Krishnamurthy A., 2008. Collective Risk Management in a Flight to Quality Episode. Journal of Finance 63, 2195-2230.

Charles, A., Darné, O., and Tripier, F., 2017. Uncertainty and the Macroeconomy: Evidence from an uncertainty composite indicator. Applied Economics (Forthcoming). Available at:https://hal-audencia.archives-ouvertes.fr/hal-01549625/document

Ciner, C., Gurdgiev, C. and Lucey, B.M., 2013. Hedges and Safe Havens: An Examination of Stocks, Bonds, Golkd, Oil, and Exchange Rates, International Review of Financial Analysis 29, 202-211.

Döpke, J., and Pierdzioch, C., 2006. Politics and the Stock Market: Evidence from Germany. European Journal of Political Economy 22, 925–943.

Doz, C., Giannone, D., and Reichlin, L., 2012. A Quasi Maximum Likelihood Approach for Large Approximate Dynamic Factor Models. The Review of Economics and Statistics, 94, 1014-1024.
31

Van Hoang, T.H., Lahiani, A., and Heller, D. 2016. Is gold a hedge against inflation? New evidence from a nonlinear ARDL approach. Economic Modelling 54(C), 54–66.

Wang, Y.S., and Chueh, Y.L., 2013. Dynamic transmission effects between the interest rate, the US dollar, and gold and crude oil prices. Economic Modelling 30 (C), 792–798.

Wang, K., Lee, Y., and Thi, T.N., 2011. Time and place where gold acts as an inflation hedge: An application of long-run and short-run threshold model. Economic Modeling 28, 806–819.

**Appendices**

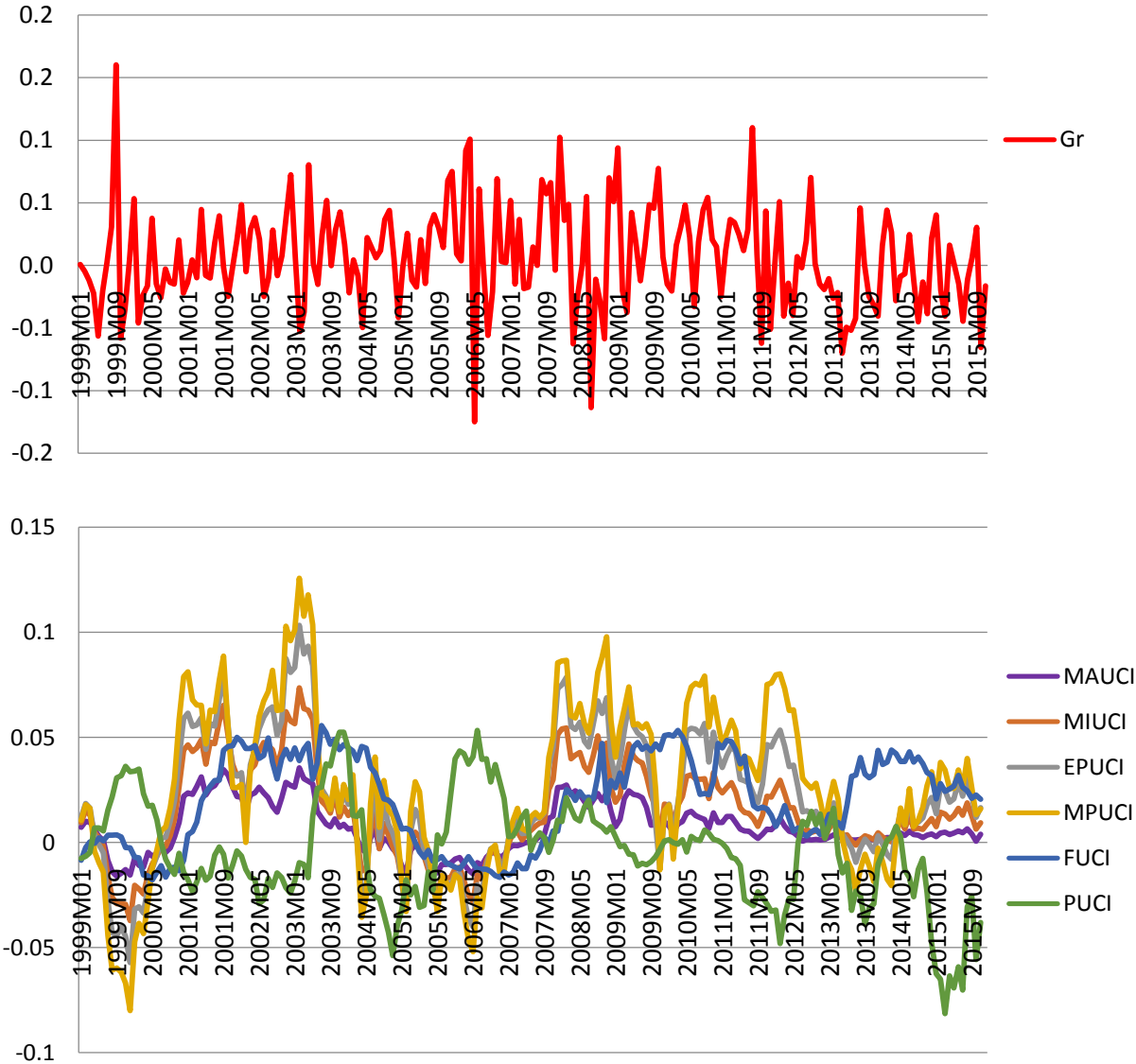

**Fig. A.1.** Data plots (the returns)